\documentclass[twocolumn,
aps,
prl,
showpacs,
amsmath,
amssymb,
floatfix,
reprint,superscriptaddress]{revtex4-1}
\pdfoutput=1

\usepackage{xcolor,mathtools,bm,mathrsfs,dsfont}
\definecolor{darkblue}{RGB}{0,0,150}
\definecolor{nightblue}{RGB}{0,0,100}

\usepackage{graphicx,mathtools,bm}
\usepackage[
colorlinks,
citecolor=darkblue,
linkcolor=darkblue,
urlcolor=nightblue]{hyperref}

\usepackage[english]{babel}
\usepackage[babel,kerning=true,spacing=true]{microtype}

\usepackage{subfigure}

\newcommand{\dd}{\mathrm{d}}

\newcommand{\beq}{\begin{equation}}
\newcommand{\eeq}{\end{equation}}

\bibpunct{[}{]}{,}{n}{}{}
\makeatletter
\def\NAT@def@citea{\def\@citea{\NAT@separator}}
\makeatother

\graphicspath{./pix/}

\begin{document}

\title{
Unified description of the classical Hall viscosity
}

\author{Tobias Holder}
\email{tobias.holder@weizmann.ac.il}
\author{Raquel Queiroz}
\email{raquel.queiroz@weizmann.ac.il}
\author{Ady Stern}
\affiliation{Department of Condensed Matter Physics,
Weizmann Institute of Science,
Rehovot 7610001, Israel}

\date{\today}

\begin{abstract}
In absence of time-reversal symmetry, viscous electron flow hosts a number of interesting phenomena, of which we focus here on the Hall viscosity. 
Taking a step beyond the hydrodynamic definition of the Hall viscosity, we derive a generalized relation between Hall viscosity and transverse electric field using a kinetic equation approach. 
We explore two different geometries where the Hall viscosity is accessible to measurement. For hydrodynamic flow of electrons in a narrow channel, we find that the viscosity may be measured by a local probe of the transverse electric field near the center of the channel. Ballistic flow, on the other hand, is dominated by boundary effects.
In a Corbino geometry viscous effects arise not from boundary friction but from the circular flow pattern of the Hall current. In this geometry we introduce a viscous Hall angle which remains well defined throughout the crossover from ballistic to hydrodynamic flow, and captures the bulk viscous response of the fluid. 
\end{abstract}

\maketitle

\paragraph{Introduction.---}
In a breakthrough insight, \citeauthor{Avron1995}~\cite{Avron1995} demonstrated the presence of a quantized observable second to the Hall conductivity in incompressible Quantum Hall states. This observable is the Hall viscosity, the antisymmetric and dissipationless part of the viscosity tensor in 2d. 
Since then, a lot of activity concentrated on working out the properties of this quantity in the gapped state~\cite{Read2009,Read2011,Son2012,Bradlyn2012,Hoyos2012,Sherafati2016}. 
Remarkably, it has been of little relevance for these studies of the Hall viscosity whether the system is assumed to be non-interacting and thus not amenable to hydrodynamic relations.
With the advent of clean materials with high mobility, attention was directed to classical electron flow in non-quantizing magnetic fields, with the hope to find a route to measure the Hall viscosity directly~\cite{Torre2015,Scaffidi2017,Falkovich2017,Pellegrino2017,Delacretaz2017,Levchenko2017}. In this case, it is necessary to restrict the discussion to viscous flow with electron-electron interactions strong enough to justify the applicability of the hydrodynamic approach. A first measurement of the Hall viscosity in Graphene was reported recently~\cite{Berdyugin2019}.

In this letter we aim to provide a unified description of the Hall viscosity which is applicable to both classical viscous fluids as well as classical non-interacting fluids. 
To this end, we examine the transverse (Hall-)response using  both a kinetic approach and the hydrodynamic constitutive relations, providing an exact mapping between the angular moments of the distribution function, transverse electric field and the Hall viscosity. 

We analyze two examples, the viscous flow in a narrow channel with diffusive boundaries and the flow through a Corbino disk with specular boundaries. 
For experiment it is desirable to relate a non-zero Hall viscosity to inhomogeneity corrections in the Hall conductivity/resistivity. In the past, this effort has been stymied by the appearance of additional correction terms of geometric or magnetic nature which depend strongly on sample details.
As we demonstrate, in classical fluids these difficulties are connected to the fact that the shear originates purely from boundary friction.
Since the Hall viscosity is a bulk quantity, its influence is most optimally measured away from contacts or sample boundaries. 
Such a bulk measurement can be realized by locally measuring Hall electric field (in a channel) or Hall current density (in a Corbino disk), both of which being quantities which can be measured in viscous fluids~\cite{Yacoby2018,Ella2018}. 
Such is possible as long as the fluid remains well in the hydrodynamic regime.
In a narrow channel we find that the bulk properties of viscous flow become inaccessible once the interaction mean free path is comparable to the channel width. The reason is intricate boundary effects related to the cyclotron motion of electrons in a magnetic field~\cite{Holder2018b}. 
To circumvent this issue we suggest a Corbino geometry which can build up shear in the flow in the absence of any boundary friction~\cite{Shavit2018}. In this case it is possible to measure by an electrical measurement the ratio of Hall viscosity to shear viscosity, a quantity which we argue to be applicable not only to classical hydrodynamic but also ballistic electron flow. Not entirely unexpected, in the ballistic case the Hall viscosity is much larger than the shear viscosity, making it likely to be detected.

\paragraph{Kinetic equation.---}
We use the standard kinetic approach with relaxation time approximation, featuring a momentum-relaxing disorder mean free path $\ell_0$ and a momentum-conserving mean free path $\ell_{ee}$ due to electron-electron interactions~\cite{deJong1995,Alekseev2016,Scaffidi2017,Alekseev2018,Holder2018b}. Given a cyclotron radius $R_c$, the effective mean free path is $\ell=(\ell_0^{-1}+\ell_{ee}^{-1})^{-1}$ and the magnetic field dependent Gurzhi length is $\ell_c=R_c\sqrt{\ell\ell_0/(4\ell^2+R_c^2)}$. The kinetic equation is
\begin{align}
\partial_{t} f
+\bm{v}\cdot\nabla_{\bm{r}} f
+e(\bm{E}+\bm{v}\times\bm{B})\cdot\nabla_{\bm{p}} f&=\mathcal{I}(f).
\label{eq:boltzmann}
\end{align}
Here, $v$ is the Fermi velocity, $e$ is the electron charge and $\mathcal{I}$ is the collision integral.
At low temperatures, it is enough to track the angular dependence $\theta$ of the distribution function along the Fermi surface.
The non-equilibrum distribution function is thus expanded as $f(\bm{r},\bm{p})=f_0-E_F(\partial_{\epsilon} f_0) h(\theta)$ with $E_F$ the Fermi energy.
With $x$ the flow direction along the channel, we denote the even (odd) angular moments of $h(\theta)$ by $h_l^{c(s)}$, with $l=0,1,...$ the angular momentum. It holds generally that the transverse electric field
\begin{align}
    \mathcal{E}_y=\tfrac{1}{R_c}h_1^c(y)
    -\tfrac{1}{\ell_0}h_1^s(y)
    +\tfrac{1}{2}\partial_y h_2^c-\partial_yh_0,\label{eq:selfconsistentEy}
\end{align} 
which is nothing else but the $\sin\theta$ component of Eq.~\eqref{eq:boltzmann}.
In this expression, $h_1^c$ is proportional to the longitudinal current density, while $h_2^c$ and $h_0$ are produced by stress. As a convention, we define the electric field $\bm{E}$ to be related to the components $\mathcal{E}_x$ and $\mathcal{E}_y$ by $\mathcal{E}_i=e\bm{E} \cdot\hat{\bm{e}_i}/E_F$, that is, $\mathcal{E}$ has units of wavenumbers. For a translationally invariant channel along $x$, the Hall current necessarily vanishes $h_1^s(y)=0$.

The Hall viscosity is defined as the coefficient of the non-dissipative part in the viscosity tensor, which reads for an isotropic system~\cite{Avron1995,Levitov2016},
\begin{align}
    T^{Hall}_{ij}&=\frac{\eta_{xy}}{2}\left(
    \epsilon_{ik}(\partial_kv_j+\partial_jv_k)
    +\epsilon_{jk}(\partial_kv_i+\partial_iv_k)
    \right).
    \label{eq:hallvisdef}
\end{align}

Equating the general definition of the viscosity tensor~\cite{Landaubook10},
\begin{align}
    T_{ij}&=\int\dd\theta h(y,\theta) v_i v_j,
\end{align}
with the definition of the Hall viscosity, Eq.~\eqref{eq:hallvisdef}, we infer for the $T_{xx}$ component that 
\begin{align}
    h_0+
    \tfrac{1}{2}h_2^c(y)&=\tfrac{1}{v} \eta_{xy}(y)\partial_yh_1^c(y).
    \label{eq:hallverbatim}
\end{align}
Note that the $T_{xx}$ component is purely non-dissipative due to absence of any flow in the $y$-direction in the channel, i.e. the  velocity gradients are nonzero only along $y$. Here, we also write out the $y$-dependences to emphasize that the Hall viscosity as introduced by Eq~\eqref{eq:hallvisdef} is not \emph{a priori} spatially independent.
The relation between $\mathcal E_y$ and $\eta_{xy}$ immediately follows,
\begin{align}
     \mathcal{E}_y&=
     \tfrac{1}{R_c}h_1^c(y)+\tfrac{1}{v}\partial_y\left[
     \eta_{xy}(y)\partial_yh_1^c(y)
     \right]
     -2\partial_yh_0(y).
     \label{eq:halleq}
\end{align}
This expression is our central result for the Hall viscosity, which holds in all discussed transport regimes, be it hydrodynamic, ohmic or ballistic. 

For a hydrodynamic state 
one expects the following relation between current and electric field 
\begin{align}
E_y&=\frac{\pi\hbar }{e^2 k_F}\left(
\frac{j_x}{R_c}+
\frac{\eta_{xy}}{v}\frac{\partial^2j_x}{\partial y^2}\right),
\label{eq:hydroEy}
\end{align}
which contains the viscous correction to the bulk Hall response.
Within a hydrodynamic approach, it is not transparent how to relate Eq.~\eqref{eq:hydroEy} with moments of the distribution function in the kinetic equation, Eq.~\eqref{eq:selfconsistentEy}. In fact, Eq.~\eqref{eq:hydroEy} follows as a special case from Eq.~\eqref{eq:selfconsistentEy} when it holds that $h_2^c\propto \partial_yh^c_1$. This happens precisely when the kinetic equation is truncated to second order ($l=2$) in angular moments, but is manifestly wrong otherwise due to the presence of higher angular moments~\cite{Holder2018b}. 
After this truncation, the Hall viscosity can be related to the mean free paths as $\eta_{xy}=v\ell_c^2\ell/2R_c\ell_0$~\cite{Alekseev2016}. The corresponding Hall resistance is for weak fields and fully diffusive walls
$\rho_{xy}\propto 1-6\ell^2/w^2$~\cite{Scaffidi2017}.
Various refinements of this result have been discussed recently~\cite{Delacretaz2017,Pellegrino2017}. Most importantly, with increasing effective mean free path, the Hall resistance suffers a sign change~\cite{Holder2018b}, indicating that the influence of the Hall viscosity entering into this expression has been overestimated.

For a complete solution of the kinetic equation one must also include electrostatic effects, which enter in the charge density $h_0$ in Eq.~\eqref{eq:halleq}. For example, in an device without backgate it is $\partial_yh_0={2/(\kappa R_c)}\partial_y\mathcal{H}(h_1^c(y))$, where $\mathcal{H}$ denotes the Hilbert transform~\cite{Holder2018b}. The electrostatic contribution remains small as long as the screening wave vector $\kappa$ fulfills $\kappa w\gg 1$ and we will not discuss it further here.

\paragraph{Measurement of $\eta_{xy}$ in narrow channels.---} In a narrow channel the current is made non-uniform by the effect of the edges, which depends on the details of the edges. Yet, the quantity we are interested in is the Hall viscosity of the bulk. Here we rely on recently developed local probes of Hall electric fields \cite{Yacoby2018,Ella2018}, to provide an experimental setup to carry out a bulk measurement.

A Taylor expansion of Eq.~\eqref{eq:halleq} to second order in $\mathcal{E}_y$ and $h_1^c$  yields 
\begin{align}
    \mathcal{E}_y(0)&=
    \frac{h_1^c(0)}{R_c}+\frac{\eta_{xy}(0)}{v}{h_1^c}''(0),
    \label{eq:Eyexp1}\\
    \mathcal{E}_y''(0)&=\frac{{h_1^c}''(0)}{R_c},
    \label{eq:Eyexp2}
    \shortintertext{and thus, restoring units}
    \eta_{xy}(0)&=\frac{v}{R_c}\left(\frac{E_y(0)}{E_y''(0)}-\frac{\pi\hbar }{e^2k_F R_c }\frac{j_x(0)}{E_y''(0) }\right).
    \label{eq:Eyexp3}
\end{align}
Here we made use of the fact that the channel center by symmetry fulfills $\partial_y\eta_{xy}=0$, even outside of the hydrodynamic regime.
This is our second important result: A prescription on how to measure $\eta_{xy}$ with a local probe~\cite{Ella2018,Yacoby2018}.
The Hall viscosity can be directly accessed by measuring the transverse Hall profile ($E_y$, $E_y''$) and the local longitudinal resistivity (which yields $j_x$) in the middle of the channel.
Remarkably, this result does not suffer from boundary corrections, a source of difficulty for alternative approaches. Instead, while the local measurement of the Hall viscosity $\eta_{xy}(0)$ at the center of the channel indeed yields the Hall viscosity, its value might not be representative of the value elsewhere if $\eta_{xy}$ acquires a spatial dependence. It is a property of hydrodynamic transport that the Hall viscosity is intrinsic and constant within the channel. Only then, Eq.~\eqref{eq:Eyexp3} is a useful estimate of the bulk Hall viscosity.  On the other hand, a non-parabolic profile of $E_y$ should be viewed a sufficient criterion for \emph{non}-hydrodynamic flow.

Let us compare this setup with the measurement of the spatial variation in the Hall resistivity \cite{Scaffidi2017,Delacretaz2017}. To leading order, the gradient expansion of the Hall resistivity reads for $w\gg R_c$
\begin{align}
    \mathcal{E}_y&\approx\rho_{xy}^{(0)}h_1^c-\rho_{xy}^{(2)} \ell_c^2\partial_y^2h_1^c.
    \label{eq:phenomenologicalEy}
\end{align}
Using the hydrodynamic approximation (truncating at $l=2$) the nontrivial part of the Hall resistance $\rho_{xy}-\rho_{xy}^{(0)}$ is
\begin{align}
   \rho_{xy}-\rho_{xy}^{(0)}&=
    \frac{2 \ell}{R_c \ell_0} 
    \left(1-\frac{w \coth (w/\ell_c)}{(1-r) \ell_c }
    \right)^{-1},
    \label{eq:fullhall}
\end{align}
where $0\leq r\leq 1$ is a measure for the specularity of the wall. The correction term to the Hall resistivity is therefore not immediately related to the Hall viscosity. 
However, we note that if we insert Eq.~\eqref{eq:fullhall} into Eq.~\eqref{eq:phenomenologicalEy},  $\rho_{xy}^{(2)}$ has a striking simplification,
\begin{align}
    \rho_{xy}^{(2)}=
    \frac{\rho_H^{(0)}h_1^c-\mathcal{E}_y}{R_c^2\partial_y^2h_1^c}
    =\frac{2 v\ell}{\ell_0 R_c}=\frac{4\eta_{xy}}{\ell_c^2},
    \label{eq:hydroexact}
\end{align}
losing its boundary dependence and indeed capturing the bulk Hall viscosity provided $\ell_c$ is  measured simultaneously.

In essence, Eq.~\eqref{eq:Eyexp3} can be taken as a general definition of the Hall viscosity in terms of measurable quantities. For hydrodynamic flow, $\eta_{xy}$ is an intrinsic constant of the fluid and thus constant along the entire channel width. In this case, it can be related to the non-local corrections of the Hall resistivity through Eq.~\eqref{eq:hydroexact}.

Let us for completeness apply this procedure to the non-interacting regime. We assume ballistic transport with diffusive boundaries in the limit $w<R_c<\ell_0$. Starting from an exact solution for the ultraballistic limit which was derived in Eqs.~(A.80,A.81) in Ref.~\cite{Holder2018b}, one can obtain an expression for $\eta_{xy}$ at the center of the channel by expanding Eq.~\eqref{eq:hallverbatim} around $y=0$,
\begin{align}
\eta_{xy}(0)\partial_yh_1^c(y)|_{y=0}&=\frac{v}{2}h_2^c(0)\\
\eta_{xy}(0)&= v w
\frac{\sqrt{2}-\log(1+\sqrt{2})}{4(2+\sqrt{2})}
\label{eq:ballisticeta}
\end{align}
For ballistic transport, $\eta_{xy}(0)$ is therefore purely extrinsic and remains finite for a small B,  but only as long as $R_c<\ell_0$. For smaller B-fields with $R_c>\ell_0$ the Hall response weakens.
We reiterate that in the ballistic case the Hall viscosity is no longer a bulk quantity but a spatially dependent function, and $\eta_{xy}(0)$ is not sufficient to determine the flow profile completely.
It is only upon increasing the magnetic field strength so that $R_c\ll w$, that it becomes possible to employ the expression for hydrodynamic transport, yielding now
$\eta_{xy}=v R_c/8$.

\paragraph{Viscous Hall angle.---}
\begin{figure}
    \centering
    \includegraphics[width=\columnwidth]{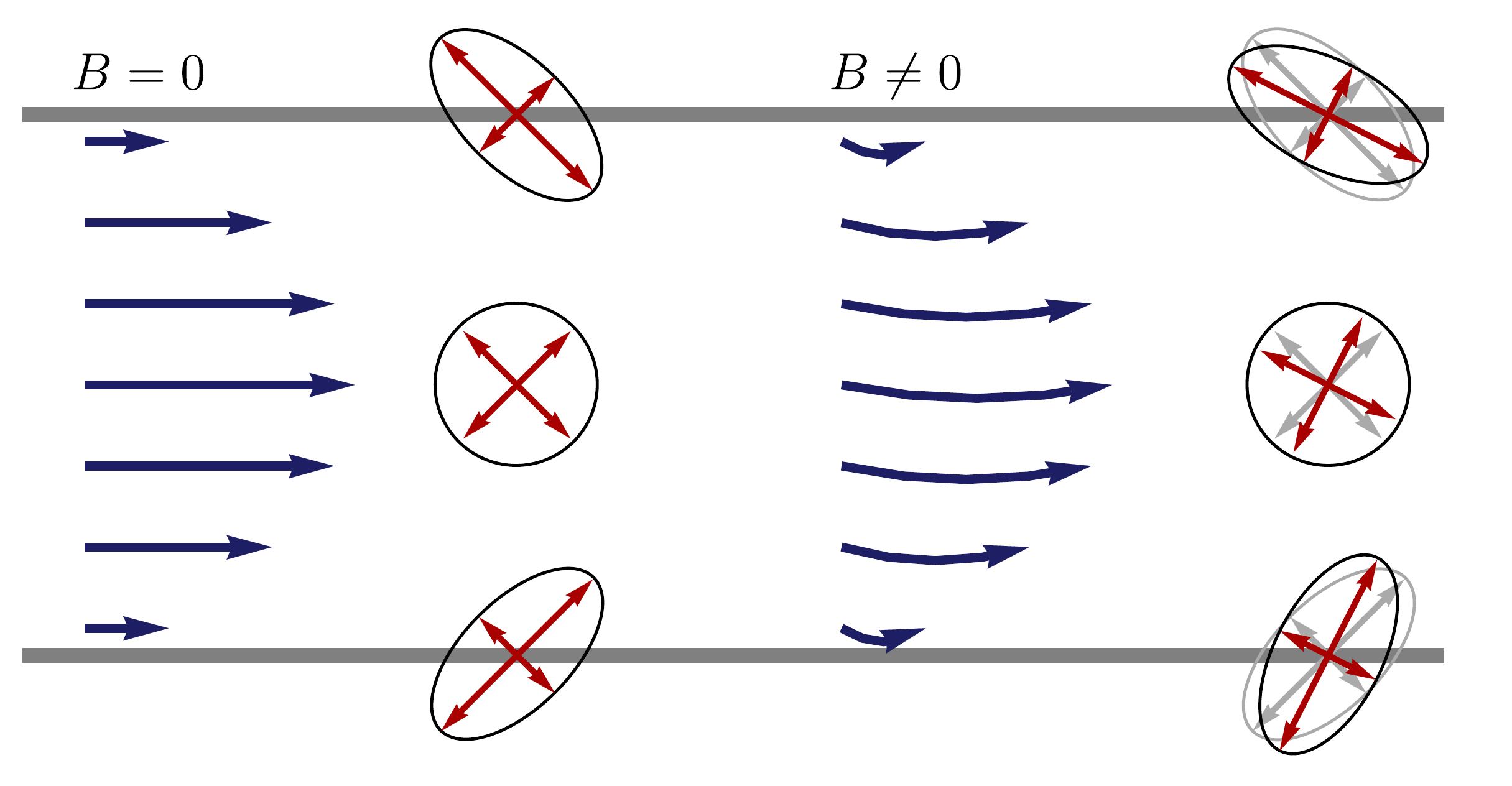}
    \caption{Rotation of the second angular moment of the distribution function in a magnetic field. Blue arrows indicate the fluid flow profile in the absence of a transverse electric field. The corresponding local distribution functions are depicted as grey circles. The red arrows are included as a guide to the eye to emphasize the angle by which the distribution function is rotated in a magnetic field.}
    \label{fig:hallangle}
\end{figure}
We relate the Hall viscosity in a narrow channel to the shear viscosity by introducing an angle that rotates the second moment of the distribution function in the presence of a magnetic field. At zero field, the second moment assumes the form depicted in Fig.~\ref{fig:hallangle}. It is antisymmetric along the channel width and encapsulates the velocity changes a flowing particle suffers upon moving from a lower speed slice to a higher speed slice in the fluid and vice-versa. While the elliptical deformation of the distribution function changes sign between lower and upper half of the channel, it is always oriented diagonally. In contrast, in the presence of a magnetic field, the orientation of the elliptical deformation is rotated by the angle $\theta_{vis}/2$ with
\begin{align}
    \tan\theta_{vis}&=\frac{\eta_{xy}}{\eta_{xx}}.
\end{align}
In the hydrodynamic limit, this viscous Hall angle becomes $\tan\theta_{vis}=\ell_{ee}/R_c$.
In the ultraballistic limit where $\ell_{ee}$, $\ell_0$ are both much larger than the system size with $\ell_{ee}\gg\ell_0$, the viscosity $\eta_{xx}$ vanishes faster than $\eta_{xy}$. The second moment is therefore dominated by the Hall viscosity and the elliptical deformation is now aligned along the axes of the channel rather than diagonal. The viscous Hall angle correspondingly approaches $\pi/2$.
This construction can be seen in close analogy to the Hall angle for conductivities, which is classically $\theta_H=\sigma_{xy}/\sigma_{xx}=\ell_0/R_c$. In both cases, the relevant scattering process which enters in the numerator is normalized by the cyclotron orbit, which decreases with increasing magnetic field until eventually the transverse component becomes quantized in the high field limit. In the presence of both momentum relaxing and momentum conserving processes the ratio $\ell/R_c$ due to both relaxation channels approaches the respective purely resistive (viscous) Hall angle for $\ell_0<w<\ell_{ee}$ ($\ell_{ee}<w<\ell_{0}$). This implies that a measurement of the viscous Hall angle in a clean but interacting system gives an estimate of both the interaction mean free path and the Hall viscosity. 

\begin{figure}
    \centering
    \includegraphics[width=.67\columnwidth]{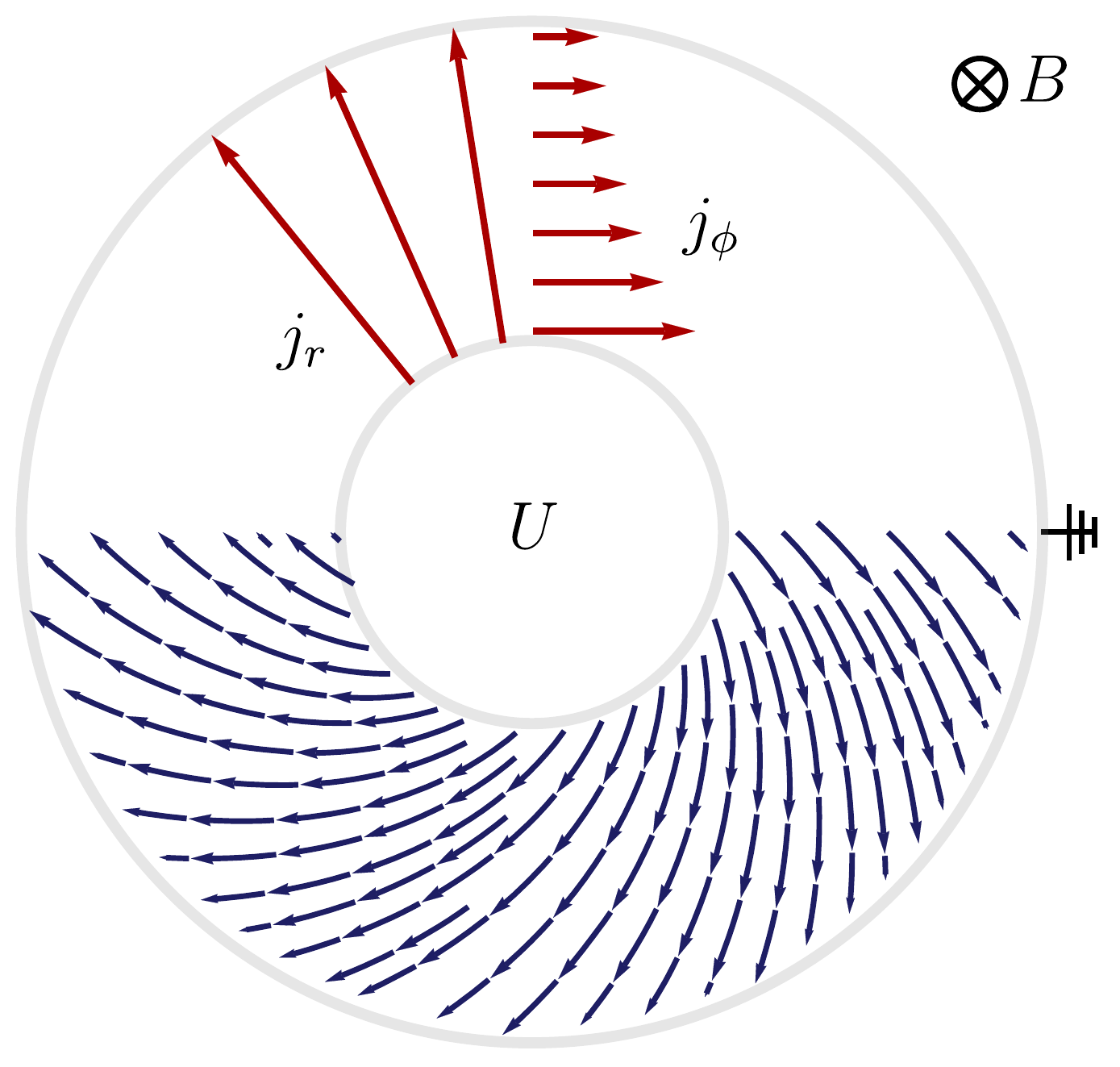}
    \caption{Viscous flow in a Corbino disk. In the presence of a magnetic field in perpendicular direction to the plane, the current between inner and outer ring (blue) has both a longitudinal and a transverse component (red). In a viscous electron fluid the Hall angle is determined by the ratio of the viscosities.}
    \label{fig:corbino}
\end{figure}

\paragraph{Measurement of $\eta_{xy}$ in a Corbino disk.---}The viscous Hall angle is particularly insightful in a Corbino geometry with specular boundaries, as depicted in Fig.~\ref{fig:corbino}. 
Since the Hall current in azimuthal direction is r-dependent due to the circular geometry, shear is present in the Hall response of the system even for fully specular boundary conditions. In contradistinction, the narrow channel must have diffusive boundaries for viscous effects to appear. 

The radial current is $j_r\propto 1/r$ due to charge conservation, while in a hydrodynamic approximation the azimuthal current $j_\varphi$ is given by
\begin{align}
    \frac{j_r}{R_c}+\frac{\eta_{xx}}{v}\nabla^2j_\varphi-
    \frac{\eta_{xy}}{v}\nabla^2j_r
    &=\frac{j_\varphi}{\ell_0}.
    \label{eq:corbino}
\end{align}
The relevant dissipative factors to compare are thus $\ell_0^{-1}$ and $\eta_{xx}/v L^2$, where $L$ is the radial size of the Corbino disk. Likewise, the two dissipationless contributions to Eq.~\eqref{eq:corbino} are of size $R_c^{-1}$ and $\eta_{xy}/v L^2$. 

Under the assumption that momentum relaxation is small ($\ell_c\gg L$)
the Hall angle becomes for the Corbino disk~\footnote{See supplementary material, where we discuss both the hydrodynamic approach and an iterative solution of the kinetic equation in a Corbino geometry.}
\begin{align}
    \tan \theta_H&
    =\frac{j_\varphi}{j_r}
    =\frac{\eta_{xy}}{\eta_{xx}}=
    \tan \theta_{vis}.
\end{align}
In other words, for a Corbino geometry with radius smaller than $\ell_c\sim\sqrt{\ell_0\ell}$ and free flow boundary conditions, the resistive Hall angle approaches the viscous Hall angle and the effects of time-reversal breaking viscous flow can be accessed directly by a measurement of the current densities. 

The considerations above neglect that the long momentum relaxing mean free path puts the flow possibly outside the applicability of the hydrodynamic formalism, as discussed before. However, it is possible to iteratively solve the full kinetic equations for the non-interacting case in powers of $1/r$. The details can be found in the supplementary material~\cite{Note1}; taking $\ell_0>L>R_c$, the viscous Hall angle in terms of microscopic quantities becomes
$\tan \theta_{vis}=2\ell_0/3R_c$. This is close to the ratio $\eta_{xy}/\eta_{xx}=\ell_0/R_c$ obtained from inserting the ballistic limit into the hydrodynamic expressions. 
This suggests that the viscous Hall angle remains a well defined bulk quantity throughout the entire crossover from hydrodynamic to ballistic flow. It is still possible that deviations appear in the intermediate regime where all length scales $\ell_0\sim\ell_{ee}\sim R_c$ are comparable.
The close connection between interaction dominated and ballistic flow in the Corbino disk compares favorably to a channel geometry, where kinetic and hydrodynamic expressions explicitly disagree in the ballistic limit. 
For a potential measurement of the Hall viscosity, we point out that the ratio $\eta_{xy}/\eta_{xx}$ is largest for ballistic transport.

\paragraph{Conclusions.---}We investigated the microscopic origin of non-dissipative viscous effects in correlated electron flow and explored the common mechanism which underlies both hydrodynamic and ballistic transport. We  advocated that the viscosity is most appropriately defined  in terms of the second angular moments of the distribution function. Most importantly, bulk hydrodynamic viscous flow appears precisely when the kinetic equation is well approximated by intrinsic viscous coefficients, that is a Hall viscosity which is spatially independent.

In ballistic flow, the effects of correlations can be analyzed employing a quantity resembling the Hall viscosity, revealing an even larger influence of viscous effects in this latter case.
The presented local definition of $\eta_{xy}$ (Eq.~\eqref{eq:Eyexp3}) makes it possible to measure the Hall viscosity in a simple channel geometry using only local electrostatic probes close to the center of the channel.

Finally, we introduce the viscous Hall angle, a quantity to capture the bulk viscous response. We show that this quantity is well defined in both ballistic and hydrodynamic limits, given that boundary friction is negligible compared to internal viscous forces. We propose that measuring this angle in a mesoscopic Corbino device allows to isolate the effects of the Hall viscosity from the intricate bulk-boundary interplay normally present in a narrow channel. 

\paragraph{Acknowledgements.---} We thank A. Rozen, J. Sulpizio, L. Ella and S. Ilani for stimulating discussions. T.H. is supported by the Minerva Foundation. A.S. is funded by the Deutsche Forschungsgemeinschaft (DFG, German Research Foundation) – Projektnummer 277101999 – TRR 183 (project B03).

\clearpage

\appendix

\setcounter{equation}{0}
\renewcommand{\theequation}{A.\arabic{equation}}

\begin{widetext}

\section{SUPPLEMENTARY MATERIAL}

In this supplementary material we solve for the azimuthal current density using the hydrodynamic approach and present the a solution of the kinetic equation for transport in a clean Corbino geometry.

\subsection{Viscous azimuthal current in a Corbino disc}
We start from Eq.~\eqref{eq:corbino} of the main text. We write it using longitudinal and transverse resistivities $\rho_{xx}$ and $\rho_{xy}$ and impose a longitudinal current $j_r=A/r$
\begin{align}
    \rho_{xy}j_r+\alpha(\eta_{xx}\nabla^2j_\varphi-
    \eta_{xy}\nabla^2j_r)
    &=\rho_{xx}j_\varphi\\
    \left(\partial_r^2+\tfrac{1}{r}\partial_r\right)j_\varphi-
    \frac{\rho_{xx}}{\alpha\eta_{xx}}j_\varphi
    &=-\frac{\rho_{xy}}{\alpha\eta_{xx}}\frac{A}{r}
    +\frac{\eta_{xy}}{\eta_{xx}}\frac{A}{r^3}.
\end{align}
Here, we used the shorthand $\alpha=\pi\hbar/e^2 v k_F$.
Rescaling the radial coordinate $r=\hat r d$ with the length scale $d=\sqrt{\alpha\eta_{xx}/\rho_{xx}}$ yields
\begin{align}
    \left(\partial_{\hat r}^2+\tfrac{1}{\hat r}\partial_{\hat r}\right)j_\varphi-
    j_\varphi
    &=-\frac{\rho_{xy}}{\rho_{xx}d}\frac{A}{{\hat r}}
    +\frac{\eta_{xy}}{\eta_{xx}d}\frac{A}{{\hat r}^3}.
\end{align}
Note that microscopically, $d^2=l_c^2/2$.
The solution to this differential equation is
\begin{align}
    j_\varphi({\hat r})&=\frac{A}{2d{\hat r}^3}
    \left(
    4\frac{\eta_{xy}}{\eta_{xx}}+(2{\hat r}^2-4)\frac{\rho_{xy}}{\rho_{xx}}
    +\pi {\hat r}
    \left(\frac{\eta_{xy}}{\eta_{xx}}-\frac{\rho_{xy}}{\rho_{xx}}\right)
    \left(2{\hat r} L_{-3}(\hat r)+({\hat r}^2+8)L_{-2}(\hat r)-{\hat r}^2I_0(\hat r)\right)
    \right)\notag\\
    &+c_1K_0({\hat r})+c_2I_0({\hat r}),
\end{align}
where $I_n(r)$, $K_n(r)$ are the modified Bessel function of first and second kind and $L_n(r)$ is the modified Struve function. $c_1$ and $c_2$ are determined by the boundary conditions. 
Under the condition that the current decays at large distances, $c_2=0$. For $\hat r\ll1$, $c_1$ enters in a subleading term, which means it only matters when the current is suppressed due to boundary friction. Otherwise, the expansion for small and large $\hat r$ yields
\begin{align}
    j_\varphi(r)&=\frac{\eta_{xy}}{\eta_{xx}}\frac{A}{r} &&r\ll d\\
    j_\varphi(r)&=\frac{\rho_{xy}}{\rho_{xx}}\frac{A}{r} &&r\gg d,
\end{align}
which is the result presented in the main text. For the expansion at large $r$ we made use of the relations
\begin{align}
    L_n(r)&=-\frac{2(n-1)}{r}L_{n-1}(r)+L_{n-2}(r)-\frac{2^{1-n}r^{n-1}}{\sqrt{\pi}\Gamma\left(n+\tfrac{1}{2}\right)}\\
    L_0(r)-I_0(r)&=-\frac{2}{r\sqrt{\pi}\Gamma\left(\tfrac{1}{2}\right)}+\mathcal{O}(r^{-2})&& \text{for $r\to\infty$}.
\end{align}

\subsection{Kinetic equation for rotationally symmetric flow}

We begin from the general expression for non-interacting transport, written with polar coordinates $(r,\phi)$ for the spatial dependence,
\begin{align}
    \cos(\theta-\phi)\partial_r h(\theta,r,\phi)
    +\frac{\sin(\theta-\phi)}{r}\partial_\phi h(\theta,r,\phi)
    +\mathcal{E}s(d/r)\cos(\theta-\phi)
    +\frac{1}{R_c}\partial_\theta h(\theta,r,\phi)&=
    -\frac{h(\theta,r,\phi)}{\ell_0}
    \label{eq:fixedframe}
\end{align}
The electric field $\mathcal{E}$ has units of wavenumber, with the physical electric field being $\bm{E}=e\mathcal{E} E_F \bm{\hat e}_r  s(d/r)$, where $s(x)$ is a dimensionless function encoding the spatial dependence.
Importantly, the problem is rotationally symmetric. This means that a rotation by $\phi$ along the disk also rotates the distribution function by this amount in momentum space. On the other hand, Eq.~\eqref{eq:fixedframe} does not implement this manifestly. We can therefore exchange the derivatives $\partial_\phi\leftrightarrow-\partial_\theta$ only after moving to the rotating frame with $\theta\rightarrow\theta+\phi$. Doing this and demanding rotational invariance for the distribution function, we are left with
\begin{align}
    \cos(\theta)\partial_r h(\theta,r)
    +\left(\frac{1}{R_c}-\frac{\sin(\theta)}{r}\right)\partial_\theta h(\theta,r)
    +\mathcal{E}s(d/r)\cos(\theta)
    &=
    -\frac{h(\theta,r)}{\ell_0}.
\end{align}
The viscous solution gives us reason to believe that the higher angular moments (in both space and momentum space here) appear at higher powers of $1/r$. Assuming that the disk is large enough that the electric field decays with increasing radius, we make the ansatz
\begin{align}
    s(x)&=\frac{1}{x}+\mathcal{O}(x^{-2})\\
    h(\theta,r)&=\sum_{n=0}^\infty g_n(\theta)\left(\tfrac{d}{r}\right)^n.
\end{align}
Sorting in powers of $r$, the tower of differential equations becomes
\begin{align}
    +\mathcal{E} d\cos(\theta)
    +\frac{g_1'(\theta)}{R_c}&=
    -\frac{g_1(\theta)}{\ell_0}&&n=1\\
    -n\cos(\theta)g_{n}(\theta)
    -\sin(\theta)g_{n}'(\theta)
    +\frac{g_{n+1}'(\theta)}{R_c}&=
    -\frac{g_{n+1}(\theta)}{\ell_0}.&&n > 1
\end{align}
Taking periodic boundary condition, $g_n(0)=g_n(2\pi)$ and $g_0=0$, the solutions are up to $n=3$
\begin{align}
    g_1(\theta)&=\mathcal{E}\ell_0
    \frac{\cos\theta+\sin\theta\ell_0/R_c}{1+\ell_0^2/R_c^2}\\
    g_{2}(\theta)&=
    c_1+\mathcal{E}\ell_0^2
    \frac{(1-2\ell_0^2/R_c^2)\cos 2\theta+3\sin 2\theta\ell_0/R_c}{(1+\ell_0^2/R_c^2)(1+4\ell_0^2/R_c^2)}\\
    g_{3}(\theta)&=
    c_2\cos\theta+c_3\sin\theta+4\mathcal{E}\ell_0^3\frac{(1-11\ell_0^2/R_c^2)\cos 3\theta+6(1-\ell_0^2/R_c^2)\sin 3\theta\ell_0/R_c}{(1+\ell_0^2/R_c^2)(1+4\ell_0^2/R_c^2)(1+9\ell_0^2/R_c^2)}
\end{align}
This confirms our suspicion, the leading $g_1$ contains only currents, while quadrupolar deformations of the Fermi surface first appear in $g_{2}$. For an arbitrary radial dependence of the electric field this conclusion remains unchanged, but additionally not exclusively $g_{2}$ but all terms $g_n$ with $n =2,4,\dots$ will contain quadrupolar contributions.
Under the assumptions mentioned in the main text, higher moments can become of similar size compared to the first moments. Of course, with increasing radius of the disk the overall decay with $r^{-n}$ suppresses the higher order angular moments carried by $g_n$ with $n>1$. Here, the relevant length to compare the radial size of the disk with is either $\ell_0$ or $R_c$, whichever is smaller.
Note that in the presence of boundary friction, this simple picture would no longer hold, as the diffusive boundary mixes and matches different angular components. 
We read off the following structure for $h_2^c$ and $h_2^s$, where the higher order terms in $r$ appear for a general radial dependence of the electric field.
\begin{align}
    h_2^c&=\mathcal{E} d \cdot f_2^c\left(\tfrac{\ell_0}{R_c}\right)\frac{\ell_0^2}{r^{2}}
    +\mathcal{O}(r^{-3})\\
    h_2^s&=\mathcal{E} d \cdot f_2^s\left(\tfrac{\ell_0}{R_c}\right)\frac{\ell_0^2}{r^{2}}
    +\mathcal{O}(r^{-3})
    \shortintertext{with the scaling functions}
    f_2^c(x)&=\frac{1-2x^2}{(1+x^2)(1+4x^2)}\\
    f_2^s(x)&=\frac{3x}{(1+x^2)(1+4x^2) }
\end{align}
Assuming that $\ell_0>d>R_c$, this simplifies to
\begin{align}
    h_2^c&=-\mathcal{E} d\frac{R_c^2}{r^{2}}\\
    h_2^s&=\tfrac{3}{2}\mathcal{E} d\frac{R_c^3}{\ell_0r^{2}}.
\end{align}

Using the definition of the viscosity, we conclude that for $\ell_0\rightarrow\infty$
\begin{align}
    \frac{\eta_{xy}}{\eta_{xx}}
    &=\left|\frac{h_2^c(r)}{h_2^s(r)}\right|
    =\frac{2\ell_0}{3R_c}.
\end{align}
In the kinetic approach, we thus find the ratio $\sigma_{xy}/\sigma_{xx}$ to be not precisely equal but very close in numbers to the ratio $\eta_{xy}/\eta_{xx}$. In the hydrodynamic formalism, both quantities converge to the same value.

\clearpage

\end{widetext}

\end{document}